\documentclass{mpe_report}

\usepackage{psfig,graphicx,epsfig}
\usepackage{color}
\usepackage{amsmath,amssymb,epic,eepic,array}

\begin{document}

\pagenumbering{arabic}
\setcounter{page}{84}

\renewcommand{\FirstPageOfPaper }{ 84}\renewcommand{\LastPageOfPaper }{ 87}

\title{Turn-over in pulsar spectra above 1 GHz}
\author{J. Kijak\inst{1}, Y. Gupta\inst{2} \and K. Krzeszowski\inst{1}}  
\institute{Institute of Astronomy, University of Zielona G\'ora,
              Lubuska 2, Zielona G\'ora, PL-65-265 Poland
\and National Centre for Radio Astrophysics, TIFR, Pune
              University Campus, Pune, India}
\maketitle

\begin{abstract}
We present the first direct evidence for turn-over in pulsar 
radio spectra at high frequencies. Two pulsars are now shown 
to have a turn-over frequency $>$ 1~GHz. We also find some 
evidence that the {\it peak frequency} of turn-over in pulsar 
spectra appears to depend on dispersion measure and pulsar age.
\end{abstract}

\section{Introduction}
A typical pulsar spectrum is steep compared to spectra of other non-thermal
radio sources and can be described by a simple power law, with a mean spectral
index of $-1.8$ (Maron et al.~\cite{maro00}).
Whereas several pulsars exhibit such a power law spectrum down to the lowest
observable frequencies, some pulsars show a low-frequency turn-over in the
spectrum (Malofeev et al.~\cite{malo94}; hereafter M94).  The frequency at which
such a~spectrum shows the {\it maximum flux} density is called the {\it peak frequency},
$\nu_{\rm peak}$, and is known to occur at $\sim$ 100 MHz (with the maximum known
about 600 MHz).  It is still an open question whether the cause of the turn-over
is some kind of absorption in the magnetosphere, efficiency loss of the
emission mechanism, or an interstellar effect (Sieber~\cite{sieb73};
hereafter S73).

In a recent paper (Kijak \& Maron~\cite{kija04}; hereafter K04), the authors
analysed the collected pulsar spectra and presented several pulsars as possibe
candidates for a~high frequency  turn-over ($\nu_{\rm peak}\sim 1$~GHz) in the
spectrum (see Fig.~1 for an example). In this paper, we present recent
observations of some of these candidates using the GMRT, and report the
evidence for $\nu_{\rm peak}$ to be around or greater than 1~GHz for a few
of them.

                 
\section{Observations and results}

The observations were made at several epochs between November 2004 and February 2005,
and some additional epochs in December 2005, using the GMRT at one of the following
three frequencies at each epoch: 325 MHz, 610 MHz or 1060 MHz, with a 16 MHz bandwidth.
We used the phased array mode with 0.512 msec sampling and 256 spectral channels
across the band (Gupta et al. \cite{gupt00}).  We observed a~total of 11 pulsars in
total intensity mode at several different epochs (see Table~1 for details).  Some
pulsars were not observed at the lower frequencies as the expected pulse broadening
due to interstellar scattering was found to be comparable or larger than the pulse
period.  To estimate the mean flux density $S_{\nu}$ of the pulsars (which represents
the total on-pulse energy, $E$, averaged over the entire pulse period, $P$,
i.e.~$S_{\nu}=E/P$), we carried out regular calibration measurements of known
continuum sources (e.g. 1311$-$22, 1822$-$096, 2350+646 and 3C48) from which an
appropriate flux density calibration scale for the pulsar data was established.

We obtained good quality pulse profiles for most of the observed pulsars.
Pulse broadening at the lower
frequencies was clearly visible in four pulsars: B1557$-$50, B1641$-$45, B1740$-$31
and B1822$-$14.  The calculated average flux densities (from
all epochs of measurements) are given in Table~1.  For each measurement, we estimated
the error in $S_{\nu}$ due to uncertainties in the calibration procedure and pulse
energy estimation. Errors due to calibration uncertainties were about 30\%.  For
multiple measurements, the error in the average $S_{\nu}$ estimate was calculated as the
mean standard deviation of the set of measurements.  There are some cases in Table~1
where we have only lower limits on the $S_{\nu}$ estimate.  This happened for cases
where the scatter broadened pulse profile was almost as broad as (or slightly broader
than) the pulsar period, making it difficult to estimate the true off-pulse power
level and hence leading to an underestimate of the total on-pulse energy.

\begin{figure}[t]
\vspace{0.7cm}
\centering
\includegraphics[width=7cm]{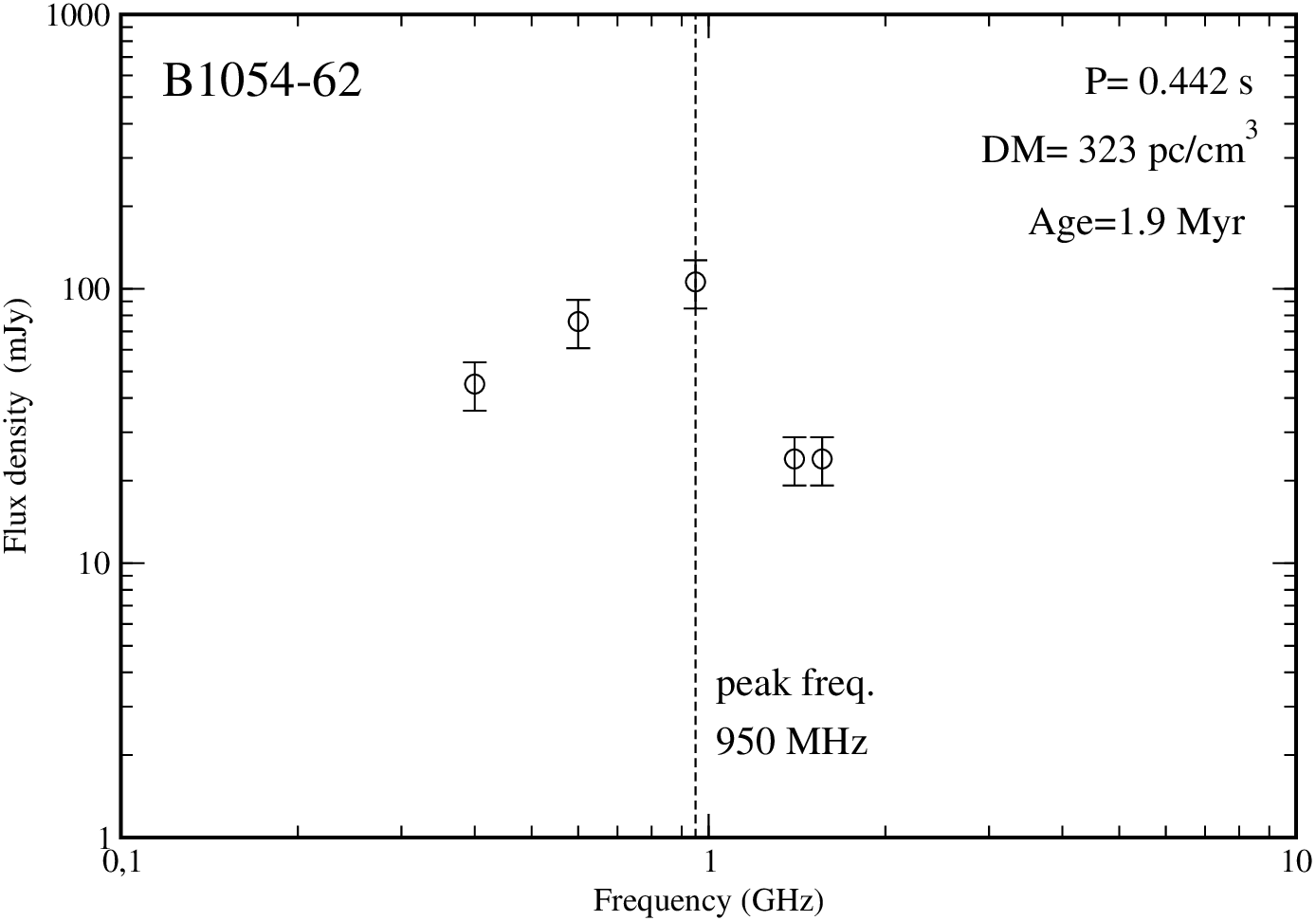}
\caption{Example of a spectrum with high-frequency turn-over. Data were
taken from C91, T93, vO97, K95 and W93 (see~References).}
\end{figure}

\begin{table}[t]
\begin{center}
\caption{Observed pulsars at the GMRT. PSRs with turn-over are marked
in bold. The total number of epochs of flux density measurements at each frequency
are given in parentheses.}
\begin{tabular}{lcrcr}
\hline\hline
& & & & \\
PSR & $DM$ & $\nu_{\rm obs}$~~ & $\left<t_{\rm obs}\right>$ &
\multicolumn{1}{c}{$\left<~S_{\rm mean}~\right>$} \\
& $\left({\rm pc~cm}^{-3}\right)$ & (MHz) & (min) & \multicolumn{1}{c}{(mJy)} \\
&  &  &  &  \\
\hline
B1557$-$50 & 261 & 325 & 16  &  128$\pm 38$ (1)  \\
           & & 610  & 9 &  66$\pm 13$ (3)  \\
           & & 1060 & 5 & 24.6$\pm 1.4$ (3) \\
\hline
B1641$-$45 & 480 & 610  & 10 & 630$\pm 53$ (3) \\
           & & 1060 & 4 & 323$\pm 52$ (2) \\
\hline
{\bf B1740$-$31} & 192 & 325  & 25 & 7.6$\pm 0.6$ (2) \\
           & & 610  & 12 & 10.6$\pm 2.6$ (3)\\
           & & 1060 & 12 & 3.0$\pm 0.3$ (4) \\
\hline
B1820$-$14 & 648 & 610  & 20 & 8.8$\pm 0.6$ (2) \\
                 & & 1060 & 13 & 1.2$\pm 0.1$ (2) \\
\hline
{\bf B1822$-$14} & 354 & 610  & 26 & 1.8$\pm 0.3$ (3) \\
           & & 1060 & 20 & 2.4$\pm 0.5$ (3) \\
\hline
{\bf B1823$-$13} & 231 & 1060 & 23 & 3.6$\pm 0.8$ (4) \\
\hline
B1830$-$08 & 411 & 610  & 12 & 18.5$\pm 3.2$ (2) \\
\hline
B1838$-$04 & 324 & 610  & 10 & 12.7$\pm 0.5$ (2) \\
           & & 1060 & 3 & 2.7$\pm 0.8$ (1)           \\
\hline
B2303+46   & 62 & 325  & 21 & 5.3$\pm 0.6$ (2) \\
           & & 610  & 18 & 2.8$\pm 1.3$ (2) \\
           & & 1060 & 16 & 1.2$\pm 0.4$ (2) \\
\hline
B2319+60   & 95 & 325  & 15 & 84.2$\pm 2.5$ (2) \\
           & & 610  & 5 & 39$\pm 12$ (1)          \\
           & & 1060 & 11 & 23.9$\pm 0.3$ (2) \\
\hline\hline
\end{tabular}
\end{center}
\end{table}

Using our results, in combination with data from the literature, we have constructed
spectra for these pulsars and find three pulsars with clear indication of a spectral
turn-over, wheras for the others we find a steadily rising
spectrum at the lower frequencies (pulsars in Table~1 which are {\it not} in bold face
font).  Below, we examine in detail the results for these 3 turn-over pulsars (along
with PSR B1054-62):

{\bf PSR B1054$-$62}. The spectrum is constructed from data published during last 15
years (see Fig.1).  This pulsar is the brightest amongst those with turn-over presented
here. The slope in the spectrum has a positive value below 1 GHz and at high frequency
the flux density drops again, showing clear evidence for $\nu_{\rm peak}$ being around
1~GHz.  The profile has a single component with a somewhat asymmetric shape at low
frequencies which is due to scatter broadening.

{\bf PSR B1740$-$31}. This pulsar was observed at three frequencies with the
GMRT, and the profile shows a single emission component that gets significantly
scatter broadened with decreasing frequency. The frequency of 610 MHz is
common with published results from Lorimer et al. \cite{lori95} (hereafter L95),
and the flux density estimates at this frequency agree quite well within the error bars. Our
$S_{\nu}$ measurement at 1060 MHz confirms the negative slope of the spectrum above
600~MHz, whereas that at 325 MHz is in agreement with the upper limit from L95.
Taken in conjunction with the small error bar on the 610 MHz flux density
estimate of L95, our results show that the spectrum has a~turn-over near 600 MHz.

{\bf PSR B1822$-$14}. The profile of this pulsar is a  single component at 1400
and 1600 MHz (\cite{goul98}), and our observations at 1060 MHz confirm
the same.  However, there is significant scatter broadening of the pulse even at
1060 MHz, which gets worse at 610 MHz.  This, coupled with somewhat low signal to
noise, makes it difficult to discern the full extent of the profile at 610 MHz.
Using the profile from the epoch with the best signal to noise, we estimate the
longitude extent of the profile and use the same width for each of the other epochs,
to estimate the  $S_{\nu}$ at 610 MHz.  From the spectrum shown in Fig. 2, it is
clear that the flux density falls at frequencies above 1.4 GHz.  Below this frequency also,
the flux density appears to be decreasing.  Though the error bars on our flux density
estimates are somewhat large, it is clear that the data points are {\it not}
consistent with a rising spectrum below 1.4 GHz.  Hence, we propose a
$\nu_{\rm peak}$ at 1.4 GHz.

{\bf PSR B1823$-$13}. The profile for this pulsar shows two emission components
which do not evolve much at the higher frequencies studied by L95 and
\cite{goul98}.  From the spectrum in Fig. 3, we conclude that the turn-over is close
to 1.6 GHz.  This is based on the very accurate flux density estimates at 1.4 and 1.6 GHz
by L95 and our own measurement at 1060 MHz which, though it has a relatively
larger error bar, has a mean value that is consistent with the L95 measurements,
for a turn-over at 1.6 GHz.

It is interesting to note that all of these pulsars have fairly high dispersion
measure ($DM$) values : $\sim$ 200 and larger. In order to investigate such effects
in the turn-over frequency in detail, we took the results for 41 pulsars with
turn-over spectra from the literature (S73; M94) and combined them with the 4 pulsars
discussed above: B1054$-$62, B1740$-$31, B1822$-$14 and B1823$-$13. This set of 45
pulsars was analyzed to check for correlations between $\nu_{\rm peak}$ and other
pulsar parameters (e.g. pulsar period, age, DM), using linear regression formulas.
Though we did not find any strong correlations, we obtained a correlation coefficient
$r\sim 0.6$ between $\nu_{\rm peak}$ and pulsar age
($\nu_{\rm peak}\propto ~\tau^{-0.23\pm 0.05}$) and DM
($\nu_{\rm peak}\propto ~DM^{0.44\pm 0.08}$), suggesting the possibility
of a correlation between these parameters and the turn-over frequency
(Figs. 4 and 5).

\begin{figure}
\vspace{0.7cm}
\centering
\includegraphics[width=8.2cm]{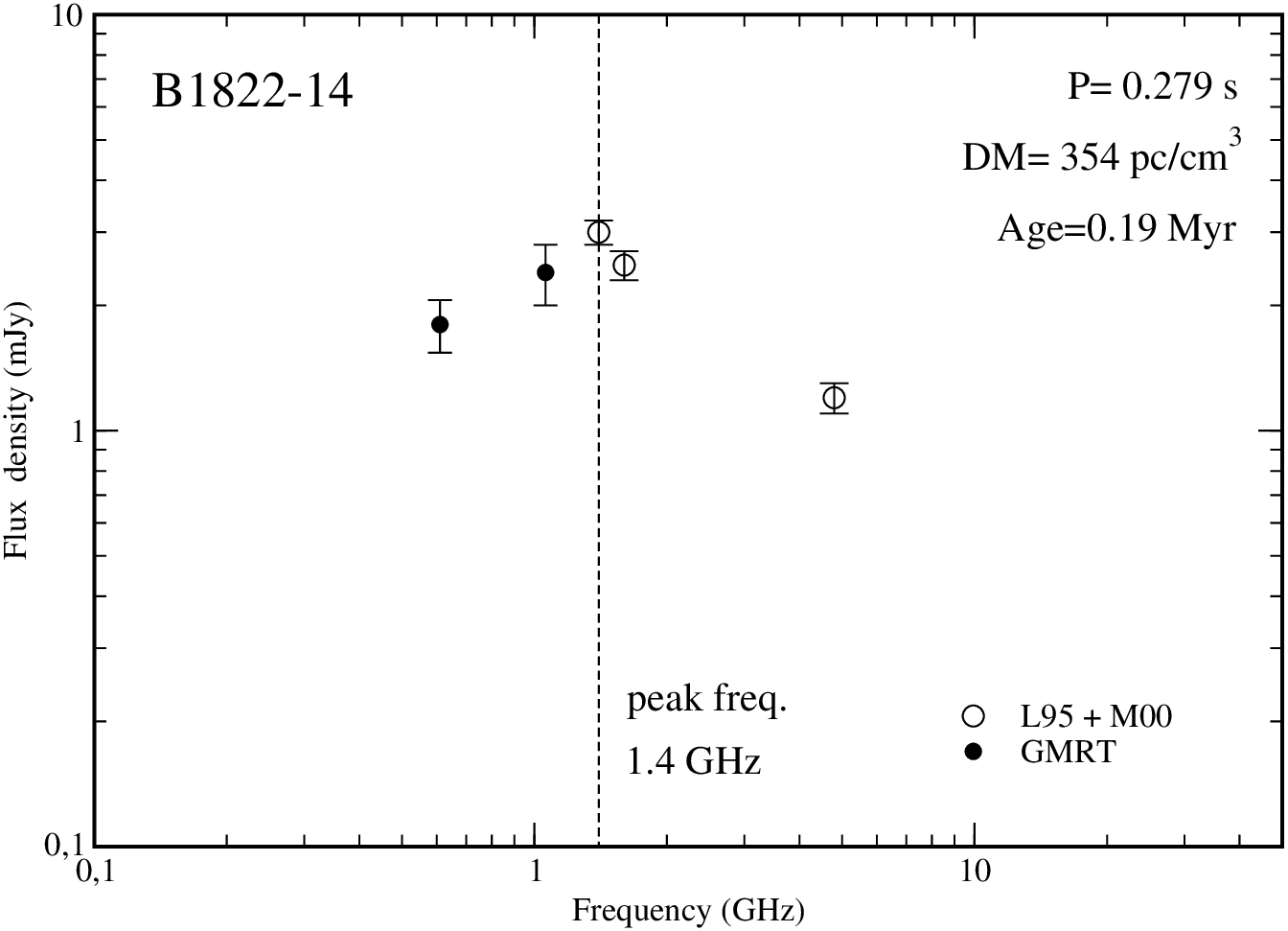}
\caption{The spectrum with turn-over above 1~GHz.}
\end{figure}

\begin{figure}
\vspace{0.75cm}
\centering
\includegraphics[width=8.3cm]{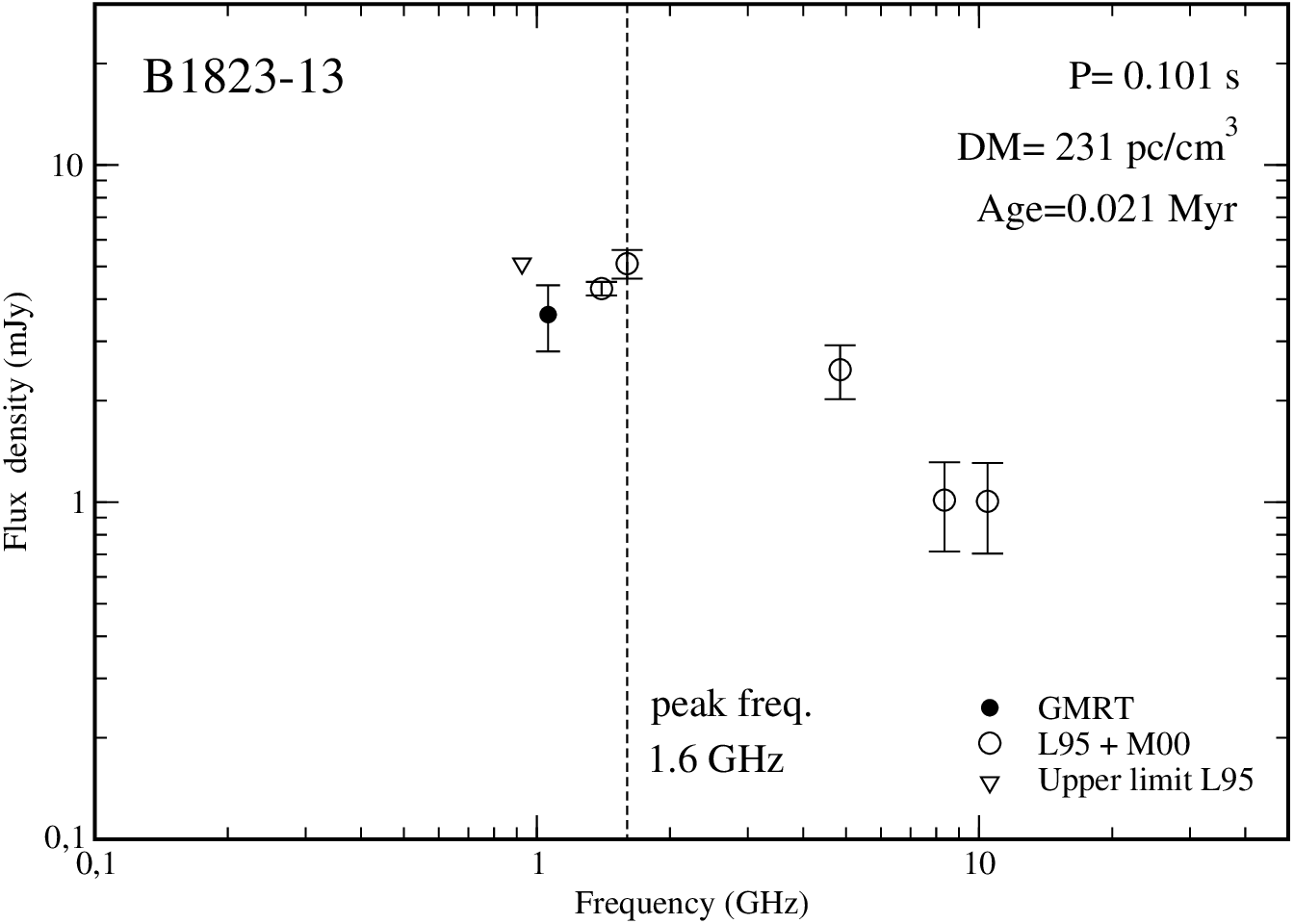}
\caption{The spectrum with turn-over above 1~GHz.}
\end{figure}

\section{Discussion and conclusions}
In their paper, K04 identified 19 pulsars as possible candidates for high frequency
turn-over spectra, based on the known $S_{\nu}$ estimates (see their Figure~1 and Table~1).
However, a careful examination of the profiles of these pulsars (using the European
Pulsar Network Data Archive\footnote{www.mpifr-bonn.mpg.de/div/pulsar/data/}) shows
that scatter broadening at the lower frequencies is comparable to or larger than
the period for some of them (B1714$-$34, B1736$-$31 and B1820$-$11). As explained
above, this can result in the flux density values being underestimated at the lower frequencies,
giving a false impression of a high-frequency turn-over.  From the low frequency
observations reported here, we can see that four more pulsars (B1557$-$50, B1641$-$45,
B1830$-$08 and B1838$-$04) turn out {\it not} to show a high frequency turn-over.
The two pulsars that we have confirmed to have a high-frequency ($>$ ~1~GHz)
turn-over are B1822$-$14 and B1823$-$13, while B1740$-$31 is shown to have turn-over
at a slightly lower frequency ($\approx$ ~600~MHz). Of the remaining pulsars in the
list of K04, there are still a few that need low frequency observations to elucidate
the true nature of their spectra (e.g. B1240$-$64, B1815$-$14, B1820$-$14, B1828$-$10,
B1823$-$11 and B1834$-$04) -- these continue to be good candidates for high-frequency
turn-over.

\begin{figure}[b]
\vspace{0.7cm}
\centering
\includegraphics[width=8.2cm]{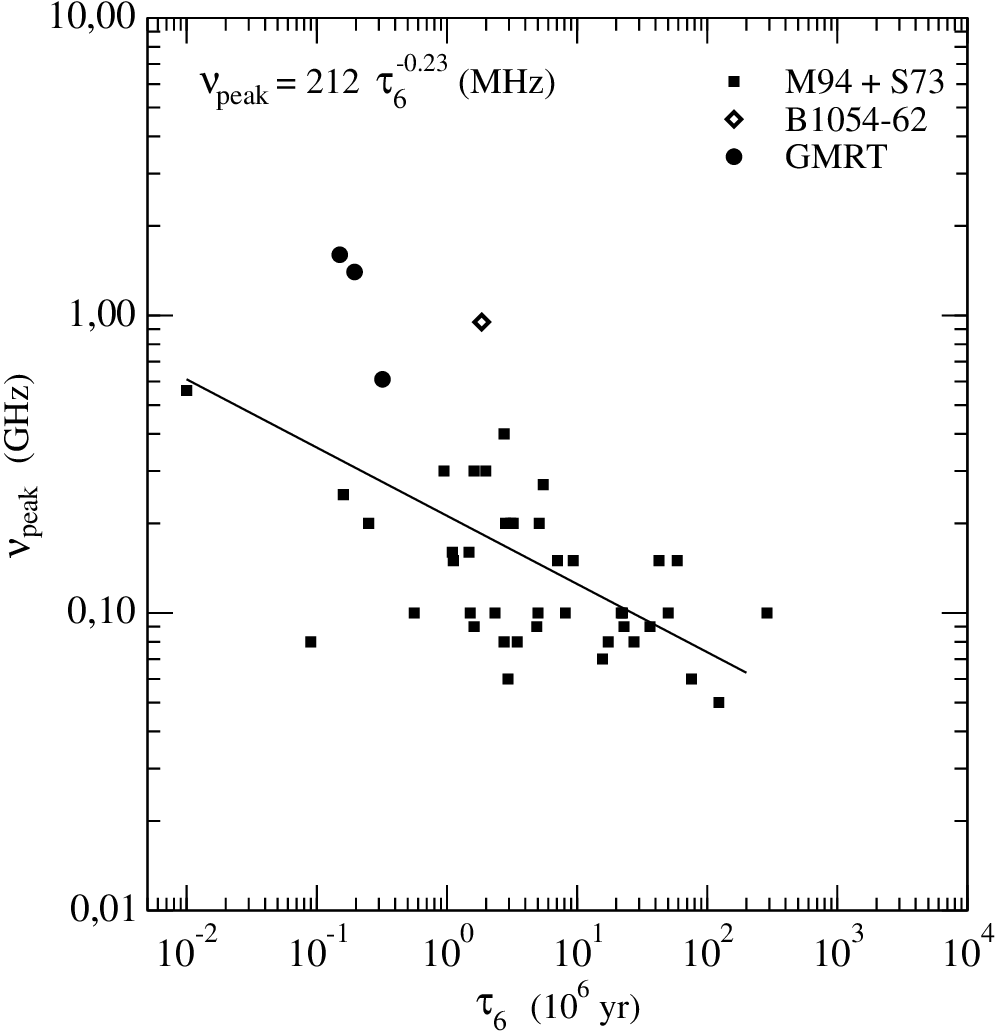}
\caption{Peak frequency versus age.}
\end{figure}

\begin{figure}[t]
\vspace{1.1cm}
\centering
\includegraphics[width=8.3cm]{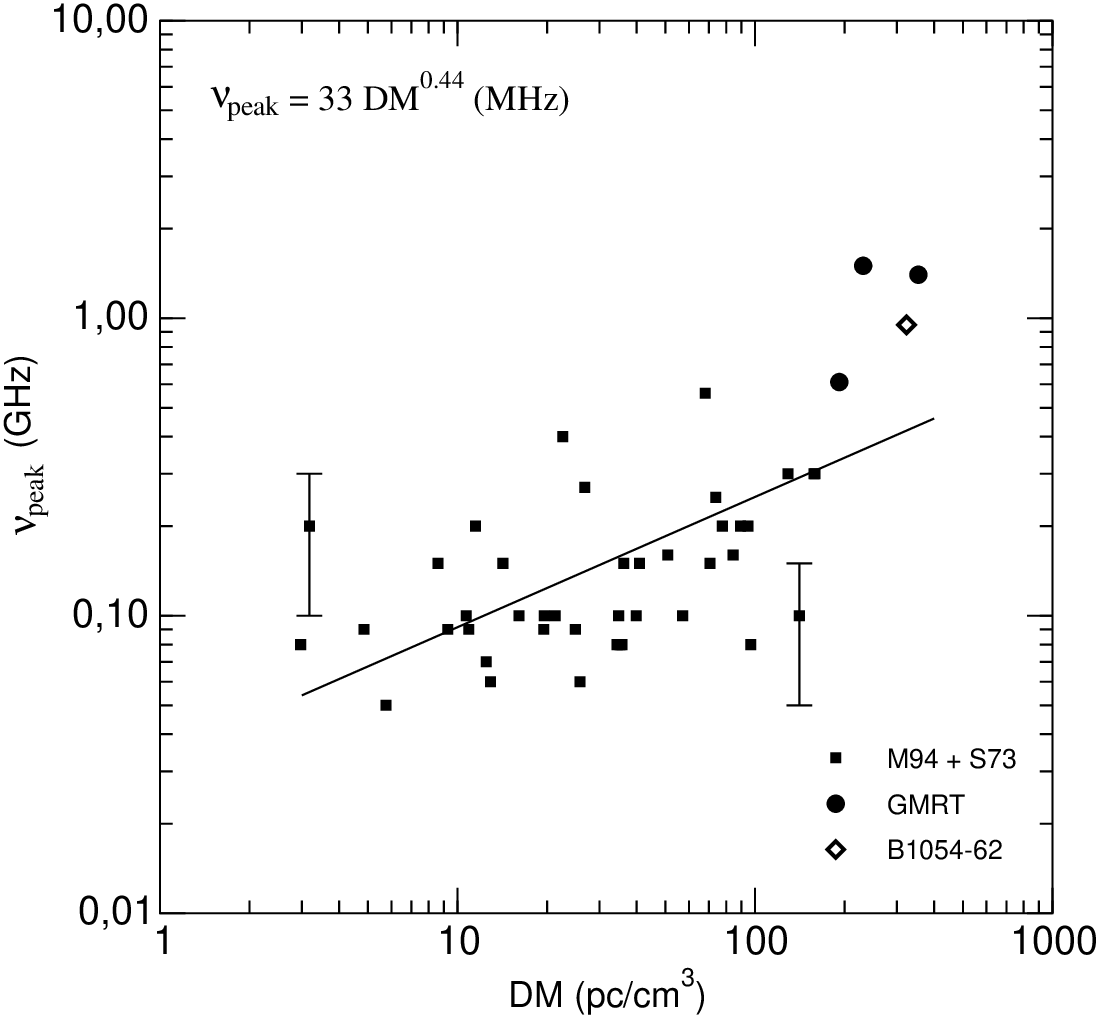}
\caption{Peak frequency versus DM.}
\end{figure}

The results thus show that there is a spread in the turn-over frequencies, with new
values reported up to and more than 1~GHz.  It is worth considering if other effects
can bias the flux density estimates at different frequencies and hence modify the observed
turn-over frequency.  One possibility is the frequency evolution of a pulsar's profile
which can cause new emission components to appear and become stronger (or old ones to
become weaker and disappear) with frequency, thereby causing a frequency dependent
variation of the total on-pulse flux density.  In the case of the pulsars studied
here, we believe that this is not a relevant possibility as the profiles are all
relatively simple and do not show significant evolution with frequency.

Another possibility is the effect of interstellar scintillations -- diffractive as
well as refractive -- that can bias the $S_{\nu}$ measured at a single epoch.  For high
DM pulsars, as the bandwidth and time scales of diffractive scintillations are
significantly smaller than the total observing bandwidth and time, the effect of
diffractive scintillations is almost completely quenched.  Though the time scale of
refractive scintillations is much larger than the observing time, the modulation index
of these is quite small and not likely to affect the $S_{\nu}$ estimates significantly.

Turning now to the statistical analysis of turn-over effects, we first note
that the correlations with other parameters can be more firmly tested if
the error bars on the low frequency $S_{\nu}$ estimates (see S73 or M94)
of the pulsars are reduced.  The large error bars translate to large
errors in the estimates of $\nu_{\rm peak}$, increasing the scatter in these
correlation studies.  Secondly, we note that it is possible that the tendency
for higher turn-over frequencies to occur for pulsars with higher DMs could
be a selection effect.  The difficulties in low frequency observations (due to
excessive pulse broadening) of high DM pulsars with low frequency turn-overs
could account for their absence (S73 and M94).  However, this would not
explain why pulsars with low DMs do not show turn-overs at around 1~GHz.
Hence, we think that this aspect needs a more detailed study.

The possible relation with pulsar age is also interesting.  Millisecond pulsars,
which are very old objects, show no evidence for spectral turn-over down to
100 MHz (Kuzmin \& Losovsky \cite{kuzm01}) and this result is consistent
with our finding.  On the other hand, our results are in contradiction with a
prospective correlation between $\nu_{\rm peak}$ and $P$. We suggest that a
period dependence of $\nu_{\rm peak}\propto P^{-0.36}$ (Malofeev \cite{malo96}),
does not exist.

Now, we turn to the question: what is the possible cause of spectral turn-over
in pulsars?  In this context, it is interesting to note that two pulsars with
turn-over at high frequencies (B1054$-$62 and B1823$-$13) have been shown to have
very interesting interstellar environments (Koribalski et al. \cite{kori95} and
Gaensler et al. \cite{gaen03}, respectively). This could suggest that the turn-over
phenomenon is associated with the enviromental conditions around the neutron stars,
rather than related intrinsically to the radio emission mechanism.  Though there
are no earlier reports of such a connection, a more detailed study on a larger
sample of pulsars is needed to address this idea more quantitatively. In this
context, future observations using GMRT (200$-$1400MHz), Effelsberg Radiotelescope
($>2$ GHz) and LOFAR ($<200$ MHz) will allow us to investigate turn-over in radio
pulsar spectra over a much wider frequency range.

Finally, we summarize our main conclusions :  First, we find clear evidence for
spectral turn-over around 1~GHz for some pulsars.  Before, the feeling in the
community was that typically the turn-over should lie around a~few hundred MHz.
Second, more and accurate pulsar flux density measurements are needed for the set of about
45 pulsars that show turn-over spectra, in order to better investigate the nature
of the {\it peak frequency} and its possible relationship with other pulsar
parameters.  Though the cause of the turn-over in pulsar spectra is still an open
question, we believe that our results can help resolve this

\vskip 0.4cm

\begin{acknowledgements}
We thank the staff of the GMRT who have made these observations possible.
The GMRT is run by the National Centre for Radio Astrophysics of the Tata
Institute of Fundamental Research.  
JK and KK acknowledge the support of the
Polish State Committe for scientific research under Grant 1 P03D 029 26.
We also gratefully acknowledge the support by the WE-Heraeus foundation.
\end{acknowledgements}



              \clearpage

\end{document}